\documentclass[reprint,amsmath,amssymb,aps]{revtex4-1}
\pdfoutput=1

\usepackage{graphicx}
\usepackage{dcolumn}
\usepackage{bm}

\begin{document}

\preprint{APS/123-QED}

\title{A novel method of searching for an axial-vector mediated force}
\thanks{}%

\author{Lei Chen}
 \altaffiliation[]{}
 \author{Jian Liu}
 \altaffiliation[]{}
\author{Ka-Di Zhu}%
 \email{zhukadi@sjtu.edu.cn}
\affiliation{Key Laboratory of Artificial Structures and Quantum Control (Ministry of
Education), School of Physics and Astronomy, Shanghai Jiao Tong
University, 800 DongChuan Road, Shanghai 200240, China,
Collaborative Innovation Center of Advanced Microstructures, Nanjing, China
}%

\date{\today}

\begin{abstract}
We present an atomic, molecular, and optical physics based method for the purpose of search for axial-vector mediated dipole-dipole interaction between electrons. In our scheme, applying a static magnetic field and a pump beam and a probe beam to a hybrid mechanical system composed of a nitrogen-vacancy  center and a cantilever resonator, we could obtain a probe absorption spectrum. Based on the study of the relationship between this spectrum and the exotic dipole-dipole interaction, we put forward our detection principle  and then provide a prospective constraint most stringent at a rough interaction range from $4\times{10}^{-8}$ to $2\times {10}^{-7}m$  . Our results indicate that this scheme could be put into consideration in relevant experimental searches .

\keywords{method, exotic dipole-dipole interaction, constraint}

\end{abstract}

\pacs{Valid PACS appear here}
\maketitle


\section{Introduction}

The role of spin in interactions between elementary particles has been a central question in physics since its discovery [1].  As we know, the concept of spin originates from the study of the anomalous Zeeman effect [2] as a consequence of the interaction of the electron's magnetic dipole moment with an external magnetic field. Due to this interaction, it is reasonable for us to think that other sorts of (the so called exotic) spin-dependent interactions might exist between fermions [1]. And many stimulated laboratory searches have been performed to study and constrain many of them [1, 3]. In these search experiments, various atomic, molecular, and optical (AMO) physics based experimental methods such as trapped ions experiments [4], nitrogen-vacancy centers in diamond [5,6], atomic spectroscopy [7], and molecular spectroscopy [8,9] etc. have been employed [10], and more and new AMO based methods are expected to be utilized in the hereafter exotic spin-dependent interactions search experiments because of the unprecedented accuracy of AMO precision measurements [1].

In this paper, we theoretically propose a new AMO based method which can be used to search for an exotic spin-dependent interaction: axial-vector dipole-dipole interaction. In our scheme, a cantilever resonator is magnetically coupled to a nitrogen-vacancy (NV) center, resulting a hybrid mechanical system[11, 12]. Then with a static magnetic field, a pump beam and a probe beam applied, a probe absorption spectrum could be obtained. Here how this absorption spectrum will be affected by the hypothetical exotic dipole-dipole interaction between electrons is studied. Based on this study, we propose our detection principle and then provide a prospective constraint. Finally , it is expected that our method could be adopted in exotic spin-dependent experimental searches.

The remainder of the paper is organized as follows: In Sec. II we present the theoretical model and derive essential dynamical equations, in Sec. III we present and discribe the numerical results, in Sec. IV we propose the detection principle and establish a prospective constraint, in Sec. V we summarize the paper and provide a outlook.

\section{Theoretical model and dynamical equations}

\begin{figure}[tbph]
\includegraphics[width=8.5cm]{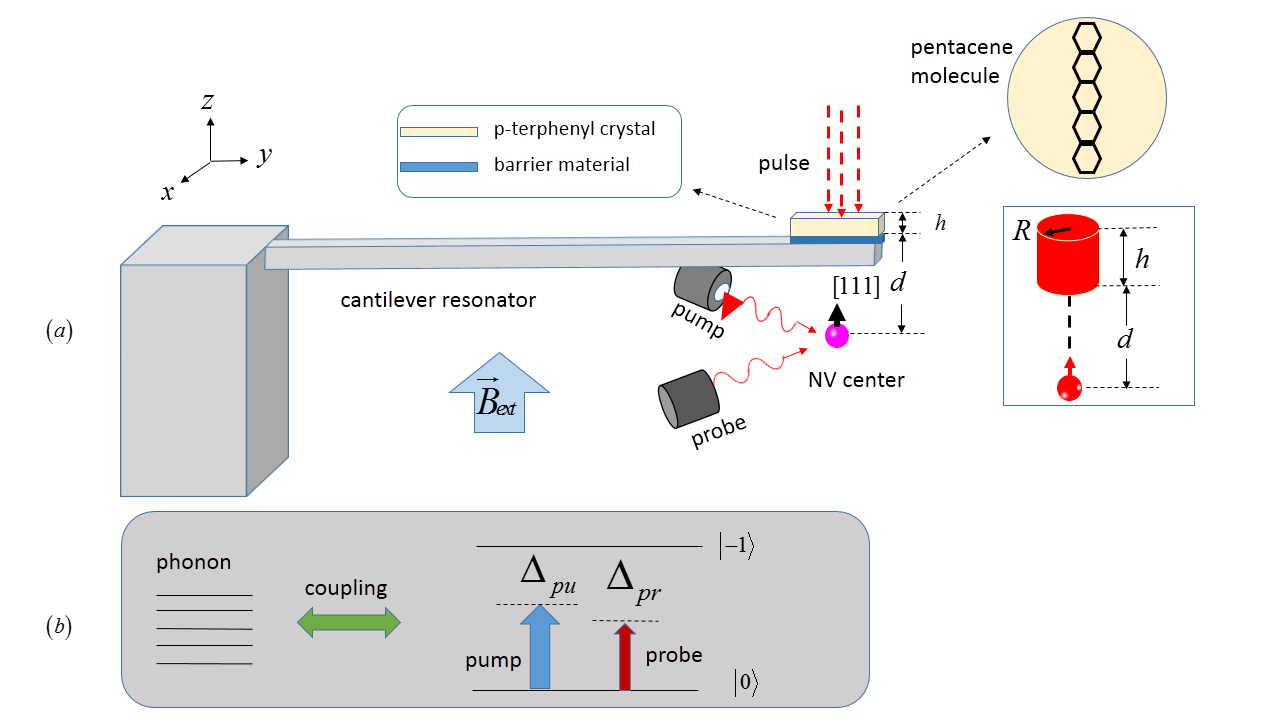}
\caption{ Setup and the energy-level diagram. (a) Schematic setup. A p-terphenyl crystal with a thickness of $h$   is placed on a cantilever resonator and there is one layer of barrier material between the two. A [111]-oriented NV center, the symmetry axis of which is assumed as z direction, is positioned at a distance $d$  under the crystal. A laser pulse is applied to the crystal, resulting a cylindrical bulk of polarized electrons with a radius of $R$  as shown in the right inset. And the symmetry axes of the bulk and NV coincide with each other. A static magnetic field  is applied along the z direction. A pump beam and a probe beam are applied to the NV center at a specific moment. (b) The energy-level diagram of the NV center spin coupled to the resonator.$ \Delta_{pu}$ and $ \Delta_{pr}$ are pump-spin detuning and probe-spin detuning respectively.}
\end{figure}
\ \ \ \ Figure 1(a) shows our setup. Here we propose a hybrid device composed of a cantilever resonator and a [111]-oriented NV center, the symmetry axis of which is assumed as z direction. A single crystal of p-terphenyl doped with pentacene-$d_{14}$ , the thickness of which is $h\approx50nm$  , is placed on the resonator. The spin density of the crystal is estimated to be $\rho=1.62\times {10}^{-3} {nm}^{-3}$ [6, 13] and the long axis of the pentacene molecule is along [111](see the upper right inset in Fig. 1(a)).  A laser pulse is applied to the crystal, resulting a cylindrical bulk of polarized electrons [6, 13] with a length of $h$ and a radius of  $R\approx500nm$ (see the right inset in Fig. 1(a)). The NV is positioned at a distance $d\approx 80nm$  under the cylinder, the symmetry axis of which coincides with the one of the NV. For the purpose of isolating the laser and fluorescence from the crystal, we place one $5nm$  thick layer of barrier material made of silver and PMMA between the single crystal and the resonator [6, 13].

Now we demonstrate the interactions between the cylindrical bulk and NV. There are two interactions between a polarized electron spin and NV: magnetic dipole-dipole interaction and axial-vector mediated interaction. The first interaction can be written
as [6]
\begin{equation}
H_{MDD}=-\frac {\mu_0 \nu^2 \hbar^2}{16 \pi r^3}[3(\overrightarrow{\sigma_p} \cdot \hat{r})(\overrightarrow{\sigma_N} \cdot \hat{r})-(\overrightarrow{\sigma_p} \cdot \overrightarrow{\sigma_N})],
\end{equation}
and the second as [6,14]
\begin{equation}
H_a = \frac {g_A^e g_A^e }{4\pi \hbar c} \frac {\hbar c}r (\overrightarrow{\sigma_p} \cdot \overrightarrow{\sigma_N}) e^{-\frac r\lambda }.
\end{equation}
Here, $\overrightarrow{\sigma_p}$  and $\overrightarrow{\sigma_N}$  are Pauli vectors of the electron spin of pentacene and that of NV respectively, $\nu$ denotes the gyromagnetic ratio of the electron spin, $\overrightarrow r$  is the displacement vector between the two electrons ,$r=|\overrightarrow r|$   and  $\hat{r}={\overrightarrow r}/r$ , $ {g_A^e g_A^e }/4\pi \hbar c$  is the dimensionless coupling constant, $\lambda=\hbar/m_b c$  is the interaction range, and $m_b$ is the mass of the boson exchanged. Because of these two interactions, NV feels an effective magnetic field from the cylindrical bulk, which can be described as [6]
\begin{equation}
B(t)=\rho P(t)\int_V B_{eff} (r,\alpha)dV,
\end{equation}
where $\rho$  is the spin density, $P(t)$  is the polarization of the bulk as a function of time,  $V$ stands for the bulk, and
\begin{equation}
B_{eff}(r,\alpha)=-\frac{\mu_0 \nu \hbar}{8\pi r^3}(3\cos^2 \alpha-1)+(\frac {g_A^e g_A^e }{4\pi \hbar c})\frac{2c}\nu \frac{e^{-\frac r\lambda }}r,
\end{equation}
$\alpha$ denotes the angle between the z direction and $\overrightarrow r$ . According to [6,13], we can assume that at some particular moment $t=t_0$ , $P(t)$  can reach its maximum value $P(t_0)=0.1$ . In the following, how the system works is interpreted.

Let us focus on the NV center first. The ground state of it is an S=1 spin triplet with three substates $|0\rangle$  and $|\pm1\rangle$ . These substates are separated by a zero-field splitting of $\omega_s/2\pi \simeq 2.88GHz$  . We apply a moderate static magnetic field ${\overrightarrow B}_{ext}$  along the z direction to remove the degeneracy of the $|\pm1\rangle$  spin states, causing the NV spin  restricted to a two-level subspace spanned by $|0\rangle$   and $|-1\rangle$ [15]. Consequently, the Hamiltonian of the NV center in the magnetic field ${\overrightarrow B}_{ext}$  can be written as $H_{NV}=\hbar \omega_s S_z$. $S_z$  together with $S^{\pm}$  characterize the spin operator.                                                                                                                                                                              Next, how the NV center is coupled to the resonator is demonstrated.

As mentioned above, NV feels an effective magnetic field $B(t)$  from the bulk of polarized electrons on the resonator. Then a magnetic coupling between NV and the resonator could be achieved through magnetic field gradient[11,16]. And the Hamiltonian of the whole system, which consists of NV, the resonator, the single crystal and the barrier material, can be described as [16]
\begin{equation}
 H_s=\hbar\omega_s S_z+\hbar\omega_r b^+ b+\hbar g(b+b^+)S_z.
\end{equation}
where $\hbar\omega_r b^+ b$  is the Hamiltonian describing the cantilever resonator, $\omega_r$  is the frequency of the fundamental bending mode and $b$  and $b^+$  are the corresponding annihilation and creation operators. Furthermore, the coupling coefficient $g$  satisfies
\begin{equation}
g=\frac{g_s \mu_B G_m a_0}\hbar,
\end{equation}
where $g_s\approx2$  , $\mu_B=9.27\times 10^{-24}A\cdot m^2$  is the Bohr magneton, $a_0$  is the amplitude of zero-point fluctuations for the whole of the resonator and the objects placed on it and it can be described as
\begin{equation}
a_0=\sqrt{\hbar/2 m \omega_r},
\end{equation}
$m$  is the mass of the whole, $G_m$  is the gradient of the magnetic field $B(t)$  at the position of the NV center. By using Eq. (3), we can derive that
\begin{equation}
G_m=\rho P(t) |u|,
\end{equation}
where $u$ is defined as
\begin{align}
u\equiv&\frac{R^2}4 \mu_0 \nu \hbar \{\frac 1{{(R^2+d^2)}^{\frac 32}}-\frac 1{{[R^2+(d+h)^2]}^{\frac 32}}\}\notag \\ &+\frac {g_A^e g_A^e}{4\pi \hbar c} \frac {4\pi \lambda c} \nu [e^{-\frac{\sqrt{R^2+d^2}}\lambda}+e^{-\frac{d+h}\lambda}\notag \\ &-e^{-\frac{\sqrt{R^2+{(d+h)}^2}}\lambda}-e^{-\frac d\lambda}].
\end{align}
Additionally, from Eqs. (6)-(8), we can derive
\begin{equation}
g=\frac {P(t) g_s \mu_B \rho}{\sqrt{2m \omega_r \hbar}} |u|.
\end{equation}

We have mentioned that $P(t)$  have a maximum value when $t=t_0$ . At this moment, we apply a pump beam and a probe beam to the NV center (see Fig. 1(a)). As a result of that, the vibration mode of the resonator can be treated as phonon mode and some nonlinear optical phenomena occur. The energy level diagram of the NV coupled to the resonator is illustrated in Fig. 1(b).  In the following we attempt to derive the first order linear optical susceptibility.

The Hamiltonian of the NV center spin in ${\overrightarrow B}_{eff}$  coupled with two optical fields reads as follows [18]:
\begin{align}
H_{int}=&-\mu(S^+ E_1 e^{-i\omega_1 t}+S^- E_1^* e^{i\omega_1 t})\notag \\
        &-\mu(S^+ E_2 e^{-i\omega_2 t}+S^- E_2^* e^{i\omega_2 t}).
\end{align}
Here $\omega_1(\omega_2)$  is the frequency of the pump field (probe field),  $E_1(E_2)$  is the slowly varying envelope of the pump field (probe field), and $\mu$ is the induced electric dipole moment. Consequently, when the magnetic field ${\overrightarrow B}_{ext}$ and two beams are applied, the Hamiltonian of the system can be described as:
\begin{align}
H=&H_s + H_{int}\notag\\
 =&\hbar \omega_s S_z +\hbar \omega_r b^+ b+\hbar g (b+b^+)S_z\notag\\
  &-\mu(S^+ E_1 e^{-i\omega_1 t}+S^- E_1^* e^{i\omega_1 t})\notag \\
  &-\mu(S^+ E_2 e^{-i\omega_2 t}+S^- E_2^* e^{i\omega_2 t}).
\end{align}
Then, we transform Eq. (12) into a rotating frame at the pump field frequency $\omega_1$  and obtain
\begin{align}
H^\prime =&\hbar\Delta_s S_z +\hbar \omega_r b^+ b +\hbar g (b+b^+)S_z -\hbar(\Omega S^+ +\Omega^* S^-)\notag\\
    &-\mu(S^+ E_2 e^{-i\delta t}+S^- E_2^* e^{i\delta t}),
\end{align}
where $\Delta_s=\omega_s-\omega_1$  , $\Omega=\mu E_1/\hbar$  is the Rabi frequency of the pump field, and $\delta=\omega_2-\omega_1$  is the pump-probe detuning.

Applying the Heisenberg equations of motion for operators $S_z , S^-$  and $\tau=b^+ +b$ , and introducing the corresponding damping and noise terms, three quantum Langevin equations can be derived as follows [17, 18]:
\begin{align}
\frac{d S_z}{dt}=&-\Upsilon_1 (S_z+\frac 12 )+i\Omega(S^+ -S^-)\notag \\ &+i{\frac \mu\hbar}(S^+ E_2 e^{-i\delta t}-S^- E_2^* e^{i\delta t}),
\end{align}
\begin{align}
\frac{d S^-}{dt}=&[-\Upsilon_2-i(\Delta_s+g\tau)]S^- -2i\Omega S_z \notag\\
&-2i{\frac \mu\hbar}S_z E_2 e^{-i\delta t}+\hat P,
\end{align}
\begin{equation}
\frac{d^2 \tau}{d t^2}+\gamma_n \frac{d\tau}{dt}+\omega^2_r \tau=-2g\omega_r S_z+\hat\kappa.
\end{equation}
 In Eqs. (14)-(16), $\Upsilon_1$  and $\Upsilon_2$  are the electron spin relaxation rate and dephasing rate respectively. $\gamma_n$  is the decay rate of the cantilever resonator.  $\hat P$ is the $\delta-correlated$  Langevin noise operator, which has zero mean $\langle \hat P \rangle=0$  and obeys the correlation function $\langle {\hat P}(t){\hat P}^+ (t^\prime)\rangle\sim\delta(t-t^\prime)$  . Thermal bath of Brownian and non-Morkovian process affects the motion of the resonator [18, 19], the quantum effect of which will be only observed when its  quality factor $Q$   satisfies $Q\gg1$  . Thus, the Brownian noise operator could be modeled as Markovian with $\gamma_n$  . The Brownian stochastic force satisfies $\langle \hat\kappa \rangle=0 $  and [19]
 \begin{equation}
\langle {\hat\kappa}^+ (t){\hat\kappa}(t^\prime)\rangle=\frac{\gamma_n}{\omega_r}\int \frac{d\omega}{2\pi} \omega e^{-i\omega(t-t^\prime)}[1+coth(\frac{\hbar \omega}{2 K_B T})].
\end{equation}

We assume  $\Upsilon_1=2\Upsilon_2$  according to [17] and could finally obtain (the detail is presented in the supplementary material)
\begin{equation}
\chi^{(1)}=\frac{\mu^2}\hbar \frac{\omega_0 A+B^* C}{C \Omega-AD},
\end{equation}
where
\begin{align}
A&=2 g^2 \eta B^* \Omega-2\Omega^2 +E(\delta+i\Upsilon_1),\notag\\
B&=\frac{\Omega\omega_0}{i\Upsilon_2 -\Delta_s +\frac{g^2 \omega_0}{\omega_r}},\notag\\
C&=2E(g^2 \eta B -\Omega), \notag\\
D&=\Delta_s -i\Upsilon_2 -g^2 \frac{\omega_0}{\omega_r}-\delta,\notag\\
E&=\Delta_s +i\Upsilon_2 -g^2 \frac{\omega_0}{\omega_r}+\delta,\notag\\
\end{align}
the auxiliary function $\eta$   satisfies
\begin{equation}
\eta=\frac{\omega_r}{\omega_r^2 -\delta^2 -i\delta\gamma_n},
\end{equation}
and the population inversion $\omega_0$  is determined by
\begin{equation}
(\omega_0 +1)[\Upsilon_2^2 +(g^2 \frac{\omega_0}{\omega_r}-\Delta_s)^2]=-2 \Omega^2 \omega_0.
\end{equation}
In addition, since it is the time $t=t_0$  when we apply two beams (see the above), all the $gs$  which appear in Eqs. (18)-(21) should satisfy
\begin{equation}
g=\frac {P(t_0) g_s \mu_B \rho}{\sqrt{2m \omega_r \hbar}} |u|,
\end{equation}
where $u$  is defined in Eq. (9).
Till now, the first order linear optical susceptibility has been derived. Then we present   numerical results in the next section.

\section{Numerical results}

We consider a Si cantilever resonator of dimensions  $(l,w,t)$ $=(10,1.2,0.05)\mu m$  with a fundamental frequency of $\omega_r \sim 2MHz$ [20], where l, w, t denote length, width, and thickness respectively. The quality factor of the cantilever resonator is assumed as $Q\sim {10}^6$ , resulting a decay rate $\gamma_n ={\omega_r}/Q =2Hz$. The mass $m$ is estimated to be $ m\sim 10^{-15} kg $. We also assume $\Delta_s =0$  , the Rabi frequency of the pump field is $\Omega = 1kHz$ , and the induced electric dipole moment is $\mu =10D$ . The NV electron spin dephasing time $T_2$  can be selected as $T_2 =1ms$ [21]. Thus the corresponding dephasing rate is $\Upsilon_2 = 1/T_2 =1kHz$  and the electron spin relaxation rate is $\Upsilon_1 =2\Upsilon_2 =2kHz$  . With these parameters, according to Eqs. (18)-(21), we plot the probe absorption spectrums ( probe absorption, i.e. ,the imaginary part of $\chi ^ {(1)} $ , as a function of pump-probe detuning $\delta$  ) around $\delta=\pm \omega_r $ for different values of $g$ in Fig. 2. Next we describe this plot in detail.
\begin{figure}[tbp]
\includegraphics[width=9cm]{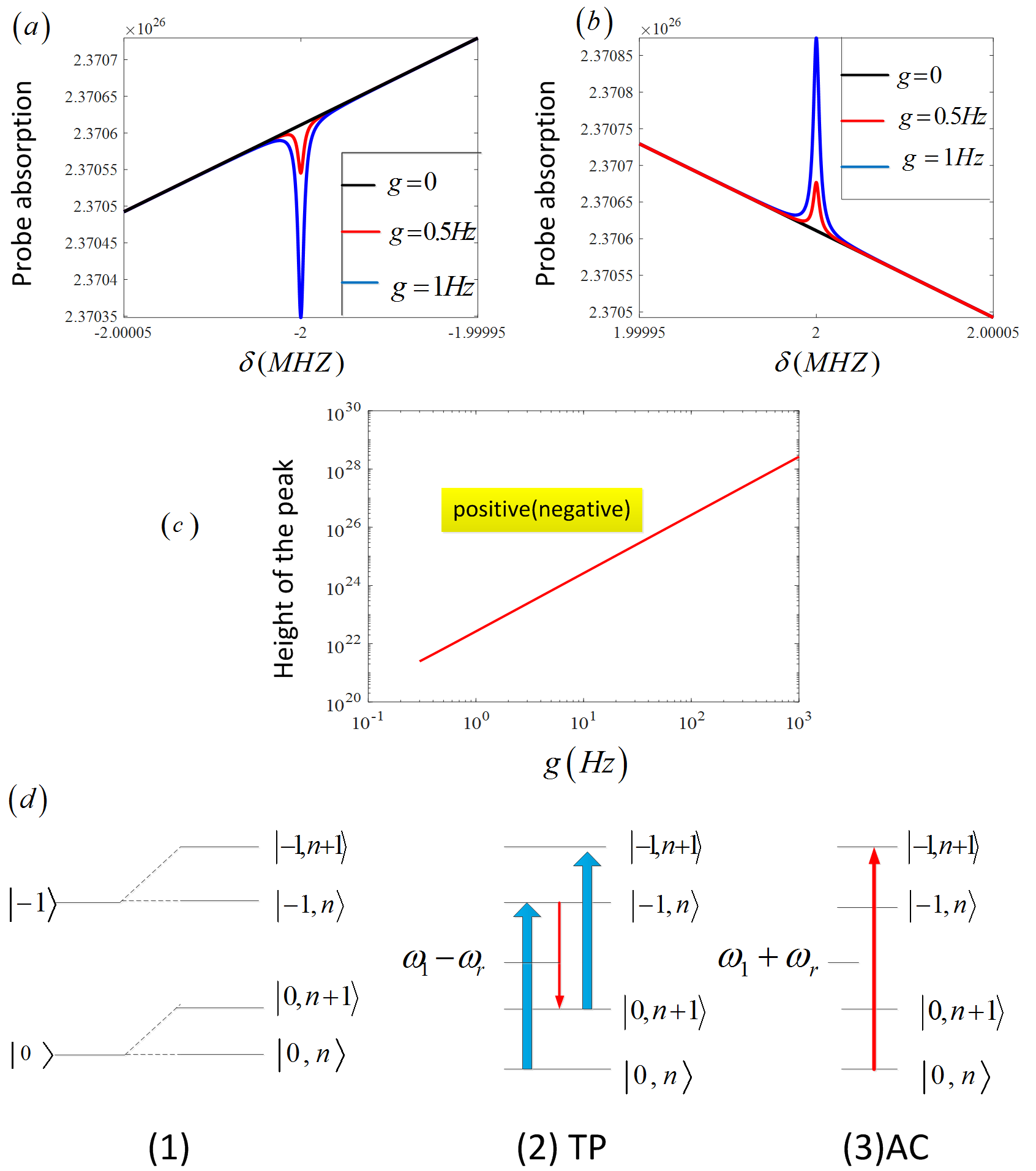}
\caption{Graph of different probe absorption spectrums around $\delta=\pm \omega_r $ ,  the relationship between height of peak and the value of $g$, and a dressed-state picture to interpret the peaks. (a)-(b) We plot three probe absorption spectrums corresponding to $g=0, 0.5, 1Hz$  respectively around
 $\delta=-\omega_r $((a)) and $\delta=\omega_r $ ((b)).  In these two subfigures, the colors of black, red, and blue are used to differentiate three values of $g$ . The parameters used are: $\omega_r \sim 2MHz$ ,$Q\sim {10}^6$ ,$\gamma_n  =2Hz$ ,$\Delta_s =0$ , $\Omega = 1KHz$,$\mu =10D$  ,$\Upsilon_1 =2\Upsilon_2 =2KHz$  . (c) The height of positive (negative) peak in the probe absorption spectrum as a function of $g$ . (d) Both the negative peaks in (a) and the positive ones in (b) are identified by the corresponding transition between dressed states of the NV electron spin. TP denotes the three-photon resonance and AC denotes the ac-Stark-shifted resonance.}
\end{figure}

In Fig. 2(a), at $-\omega_r-50Hz\leq\delta \leq -\omega_r+50Hz$ we plot three probe absorption spectrums corresponding to $g=0, 0.5Hz, 1Hz$  respectively. In the case of $g=0$  (black), there is only a straight line. When the value of $g$  grows to $0.5Hz$  (red), a negative sharp peak centered at $\delta=-\omega_r $  appears in the corresponding probe absorption spectrum, the rest of which coincides with the black straight line. When the value of $g$ continues grows to $1Hz$ (blue), the corresponding probe absorption spectrum is the same as the case of $g=0.5Hz$ , except the negative peak is larger. In Fig. 2(b), we plot the same three probe absorption spectrums at $\omega_r-50Hz\leq \delta \leq \omega_r+50Hz$ . There is only a black straight line in the case of $g=0$ . And a positive steep peak centered at $\delta=\omega_r $  appears for the cases of the red curve ($g=0.5Hz$ ) and the blue ( $g=1Hz$), while the rest of these two curves coincides with the straight line. In addition, the blue peak is larger than the red, which is the same as the situation in the Fig. 2(a). We can also take other values of $g$ into consideration, though not illustrated in Fig. 2. Then We summarize the probe absorption spectrums around $\delta=\pm \omega_r $, i.e. $\delta \in [-\omega_r -50Hz , -\omega_r +50Hz]\cup [\omega_r -50Hz , \omega_r +50Hz]$. For the probe absorption spectrum of $g=0$ , there is no peak at $\delta=\pm \omega_r $  , around each of which is only a straight line. On the contrary, when the value of $g$  grows to $0.3 Hz$  from $0$ , two evident peaks including a negative and a positive one appear in the corresponding probe absorption spectrum. Here the negative peak is centered at $\delta=-\omega_r $ while the positive one at $\delta=\omega_r $. And the rest of the spectrum coincides with the case of $g=0$. If the value of $g$  continues to grow  from $0.3Hz$ but not exceed $1kHz$ , both two peaks centered at $\delta=\pm \omega_r $ would keep becoming larger while the rest of the corresponding spectrums will coincide with the case of $g=0$. .Furthermore, It is found that the heights of  the two peaks in the same probe absorption spectrum corresponding to $0.3Hz\leq g \leq 1kHz$ are equal to each other. The height of the positive(negative) peak can be considered  as a function of $g$ whose domain is $0.3Hz\leq g \leq 1kHz$. And the graph of this function is plotted in Fig. 2(c).

Two peaks in a probe absorption spectrum for any positive value of $g$ can be interpreted by a dressed-state picture, in which the original energy levels of the NV electron spin $|-1\rangle$  and $|0\rangle$  have been dressed by the phonon mode of the cantilever resonator. Consequently, $|-1\rangle$  and $|0\rangle$  split into dressed states $|-1,n\rangle$  and $|0,n\rangle$  , where $|n\rangle$   denotes the number states of the phonon mode (see part (1) of Fig. 2(d)). The feature of the negative peak can be interpreted by TP, which denotes the three-photon resonance. Here the NV spin makes a transition from the lowest dressed state $|0,n\rangle$   to the highest dressed state $|-1,n+1\rangle$   by the simultaneous absorption of two pump photons and the emission of a photon at $\omega_1 -\omega_r$ (see part (2) of Fig. 2(d)). Meanwhile, the feature of the positive peak corresponds to the usual absorptive resonance of the NV spin as modified by the ac Stark effect, shown by part (3) of Fig. 2(d).

Till now, the relationship between the probe absorption spectrum around $\delta=\pm \omega_r $ and the value of $g$ has been demonstrated. Besides, the relationship between $g$ and the coupling constant ${g_A^e g_A^e}/{4\pi \hbar c}$ has been made clear through Eq. (22). Based on these two points, we propose our principle of detection of the axial-vector mediated dipole-dipole interaction and then a prospective constraint in the following.
\section{Detection principle and prospective constraint}

Our detection procedure consists of two steps. Firstly, we perform the experiment based on our model and then obtain a probe absorption spectrum. Then through this spectrum around $\delta=\pm\omega_r$ we estimate or constrain the value of $g$. Secondly, We specify the interaction range $\lambda$ . Then according to the relationship between $g$ and ${g_A^e g_A^e}/{4\pi \hbar c}$(Eq. (22)), we determine the value of ${g_A^e g_A^e}/{4\pi \hbar c}$ or establish an upper bound of it. In order to perform the second step well, we analyze Eq. (22) in the following.

The equation (22) is rewritten here:
 \begin{equation*}
g=\frac {P(t_0) g_s \mu_B \rho}{\sqrt{2m \omega_r \hbar}} |u|,
\end{equation*}
where
\begin{align*}
u=&\frac{R^2}4 \mu_0 \nu \hbar \{\frac 1{{(R^2+d^2)}^{\frac 32}}-\frac 1{{[R^2+(d+h)^2]}^{\frac 32}}\} \\ &+\frac {g_A^e g_A^e}{4\pi \hbar c} \frac {4\pi \lambda c} \nu [e^{-\frac{\sqrt{R^2+d^2}}\lambda}+e^{-\frac{d+h}\lambda} \\ &-e^{-\frac{\sqrt{R^2+{(d+h)}^2}}\lambda}-e^{-\frac d\lambda}].
\end{align*}
According to the previous sections, the values of parameters $P(t_0 ),g_s ,$ $\mu_B ,\rho, m, \omega_r ,\hbar, R, \mu_0 ,\nu, d, h, \lambda, c$ in Eq. (22) are fixed. Then using this equation and through calculation, with the assumptions of $g\ge {10}^{-2}Hz$ and ${10}^{-10}m\leq \lambda \leq {10}^{-1}m$, we can derive that:
\begin{align}
g=-\frac {4\pi\lambda cP(t_0)g_s \mu_B \rho u_1}{\nu\sqrt{2m \omega_r \hbar}}\frac {g_A^e g_A^e}{4\pi \hbar c}-\frac {P(t_0)g_s \mu_B \rho R^2 \mu_0 \nu \hbar u_2}{4\sqrt{2m \omega_r \hbar}},
\end{align}
and
\begin{align}
\frac {g_A^e g_A^e}{4\pi \hbar c}=-\frac{\nu\sqrt{2m \omega_r \hbar}}{4\pi\lambda cP(t_0)g_s \mu_B \rho u_1}g-\frac{R^2 \mu_0 \nu^2 \hbar u_2}{16\pi\lambda c u_1},
\end{align}
where $u_1$ and $u_2$ are defined as
\begin{equation*}
u_1\equiv e^{-\frac{\sqrt{R^2+d^2}}\lambda}+e^{-\frac{d+h}\lambda} -e^{-\frac{\sqrt{R^2+{(d+h)}^2}}\lambda}-e^{-\frac d\lambda}
\end{equation*}
and
\begin{equation*}
u_2 \equiv\frac 1{{(R^2+d^2)}^{\frac 32}}-\frac 1{{[R^2+(d+h)^2]}^{\frac 32}}
\end{equation*}
respectively.

\begin{figure}[tbp]
\includegraphics[width=8.5cm]{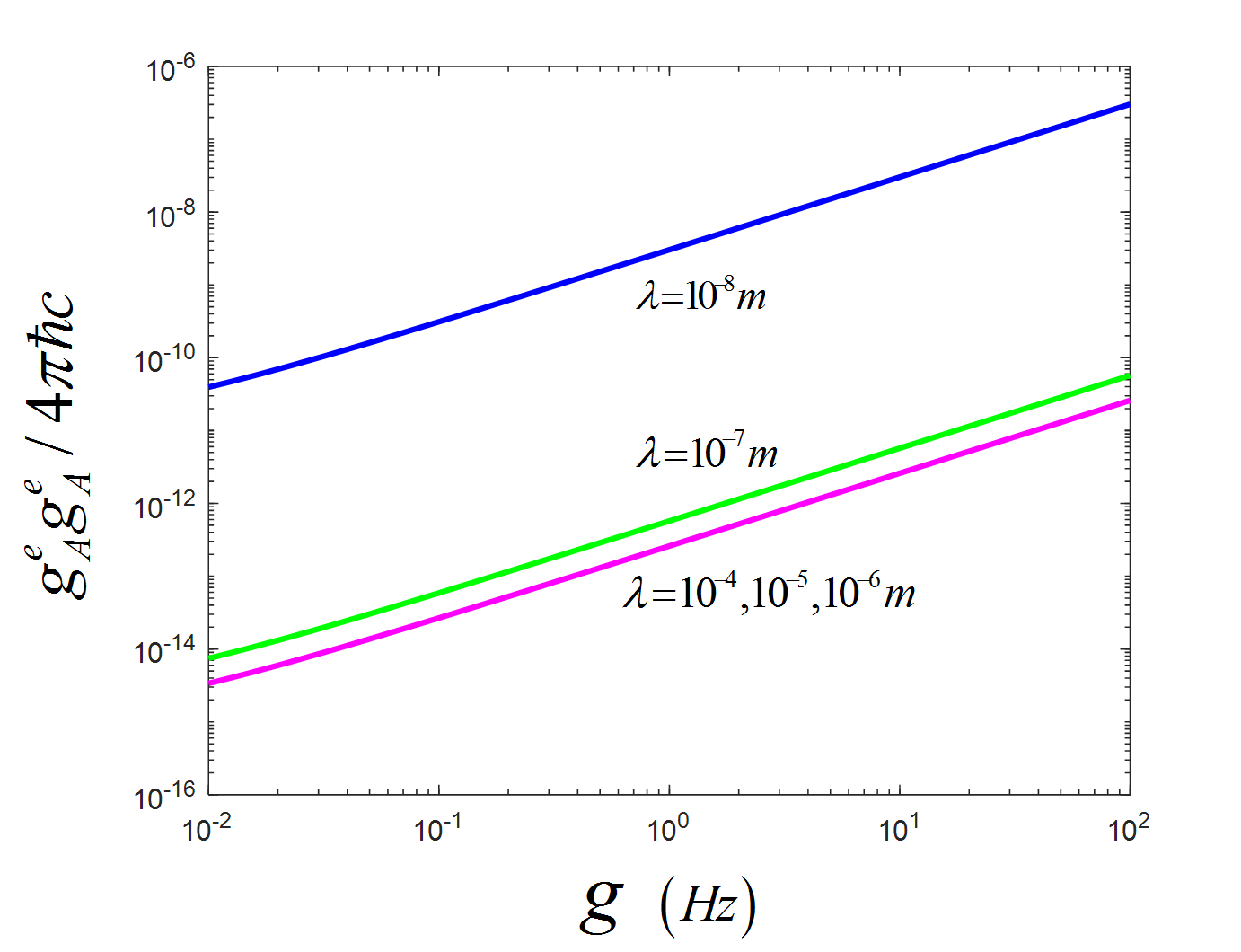}
\caption{Graphs of functions. We assume $\lambda$ takes five values of ${10}^{-n}m$ where $n=4,5,6,7,8$. Consequently, according to Eq. (24), $ {g_A^e g_A^e}/{4\pi \hbar c}$ can be considered as five functions of $g$ with ${10}^{-2}Hz\leq g \leq {10}^{2}Hz$. We plot the graphs of them with different colors. The blue and green lines correspond to the cases of $\lambda={10}^{-8}m,{10}^{-7}m$ respectively, while the carmine line simultaneously denotes the cases of $\lambda={10}^{-4}m,{10}^{-5}m, {10}^{-6}m$ .}
\end{figure}
Now focus on Eq. (24). According to it, $ {g_A^e g_A^e}/{4\pi \hbar c}$ could be considered as a function of $g$ if $\lambda$ is reasonable specified and $g$ satisfies $g\ge {10}^{-2}Hz$. Then if $\lambda$ takes values of ${10}^{-n}m$ where $n=4,5,6,7,8$ respectively, five relating functions with ${10}^{-2}Hz\leq g \leq {10}^{2}Hz$ would be obtained. And the graphs of them are plotted in Fig. (3). Then, Several findings of this figure denoted by (I)-(III) are listed: (I) The graphs corresponding to $\lambda=10^{-n} m$ where $n=4,5,6,$ overlap completely. (II) One value of $g$ possibly corresponds to different values of $ {g_A^e g_A^e}/{4\pi \hbar c}$  when $\lambda$ takes different values. (III) For any one of these five graphs , the value of $ {g_A^e g_A^e}/{4\pi \hbar c}$  keeps decreasing as the value of $g$ decreases continuously. Furthermore, the finding (III) can be generalized to a easily proved conclusion: If we assume ${10}^{-10}m\leq \lambda \leq {10}^{-1}m$ and $g\ge {10}^{-2}Hz$ , a smaller value of $g$ would correspond to a smaller value of $ {g_A^e g_A^e}/{4\pi \hbar c}$. Till now the demonstration of our detection method has just been finished. Then according to this method, we set a prospective constraint for ${g_A^e g_A^e}/{4\pi \hbar c}$  in the following.

We assume that an evident peak could be observed in a probe absorption spectrum. Based on this and the numerical result that there are two evident peaks centered at $\delta=\pm \omega_r $ respectively in the absorption spectrum corresponding to $g=0.3Hz$, we assume that the minimum value of $g$ could be identified is $0.3Hz$. We also assume that in the relating experiment the exotic dipole-dipole interaction would not be observed if $\lambda$ is assumed as
${10}^{-10}m\leq \lambda \leq {10}^{-1}m$. Combining  the second assumption and the third one we can conclude that the value of $g$ corresponding to the experimentally generated probe absorption spectrum would satisfy
\begin{equation}
g<g_c,
\end{equation}
where $g_c \equiv 0.3Hz$.  Using (25) and Eq. (22), we can obtain
\begin{equation}
{g_A^e g_A^e}/{4\pi \hbar c}<-\frac{g_c \sqrt{2m \omega_r \hbar} \nu+P(t_0)g_s \mu_B \rho u_2 {\frac{R^2}4}\mu_0 \nu^2 \hbar}{4\pi\lambda u_1 P(t_0)g_s \mu_B \rho c},
\end{equation}
where $\lambda$ satisfies ${10}^{-10}m\leq \lambda \leq {10}^{-1}m$ and $u_1$ and $u_2$ have been defined in the above. And this inequation sets upper limits on ${g_A^e g_A^e}/{4\pi \hbar c}$ as a function of the interaction range $\lambda$ the domain of which is $[10^{-10}m, 10^{-1}m]$.

\begin{figure}[tbp]
\includegraphics[width=8.5cm]{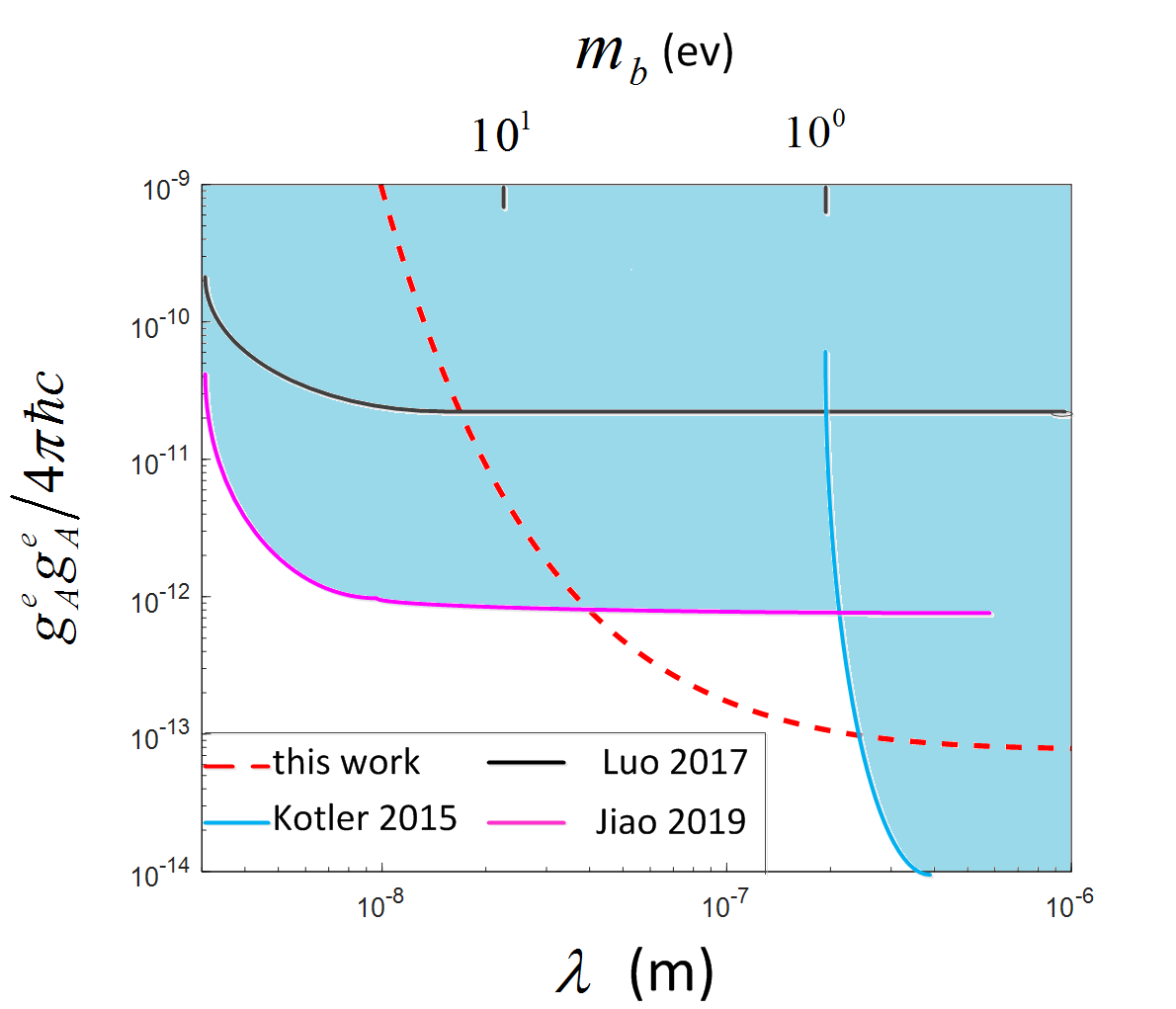}
\caption{Upper limits on the dimensionless coupling constant ${g_A^e g_A^e}/4\pi \hbar c$  as a function of the interaction range $\lambda$ and mass of the exchanged boson $m_b$ . The red curve represents our work, while the black, blue and pink ones correspond to the results from Refs. 22, 4 and 23 respectively. The dark green region is  excluded parameter space. }
\end{figure}
Up to now we have established a prospective constraint for the coupling constant ${g_A^e g_A^e}/{4\pi \hbar c}$, part of which is illustrated in Fig. 4. Then we focus on the figure. Kotler et al. established an constraint at the micrometer scale  based on a measurement of the magnetic interaction between two trapped $ ^{88}{Sr}^+$  ions [4]. Luo et al. obtained a constraint by analyzing a measurement of the magnetic dipole-dipole interaction between two iron atoms [22]. Jiao et al. set an constraint at the nanometer scale with a type of molecular rulers [23]. The dark green region is excluded parameter space. It is found that our constraint is most stringent at a rough interaction range from $4\times {10}^{-8}$   to $2\times {10}^{-7}m$ .  Of course, our constraint is only an estimated one and not accurate, and the achievement of the real or right constraint needs relevant experimental search and more theoretical analysis or calculation.

\section{Summary and outlook}

In this paper we have developed an AMO based method which could be utilized to detect axial-vector dipole-dipole interaction between polarized electrons. In our scheme, with two beams and a static magnetic field applied to a spin-resonator hybrid mechanical system, we could obtain a probe absorption spectrum through which we determine the exotic dipole-dipole interaction exists or not. Furthermore, we provide a prospective constraint,
which is most stringent at about  $4\times {10}^{-8}m<\lambda <2\times{10}^{-7}m$ . Certainly, we hope our method could be realized experimentally in the near future. Besides, in my other work [24], a similar scheme, in which also a NV based hybrid spin-resonator system is employed and two beams are applied, is proposed in order to detect exotic monopole-dipole interaction. Finally, we hope that work combined with the present one can be enlightening in the area of experimental searches for exotic spin-dependent interactions.

\begin{acknowledgments}
This work was supported by National Nature Science Foundation of China (11274230.11574206).
\end{acknowledgments}

\end{document}



\section*{The derivation of first order linear optical susceptibility}
To go beyond weak coupling, we can always rewrite each Heisenberg operator as the sum of its steady-state mean value and a small fluctuation with zero mean value as follows:
\begin{equation}
S^- =S_0^- +\delta S^- ,S_z =S_0^z +\delta S_z ,\tau=\tau_0 +\delta \tau.
\end{equation}
Next, we insert these equations into the Langevin equations (14)-(16), neglecting the nonlinear term $\delta \tau \delta S^-$  . Since the optical drives are weak, all operators could be identified with their expectation values and the quantum and thermal noise terms could be dropped [1]. As a consequence, the linearized Langevin equations can be written as:
\begin{align}
\langle\delta {\dot S}_z \rangle=&-\Upsilon_1 \langle\delta  S_z \rangle +i\Omega \langle\delta {(S^-)}^* \rangle-i\Omega^* \langle\delta S^-\rangle \notag\\ &+\frac{i\mu E_2 e^{-i\delta t}} \hbar \langle\delta {(S^-)}^* \rangle - \frac{i\mu E_2^* e^{i\delta t}} \hbar \langle\delta S^- \rangle,
\end{align}
\begin{align}
\langle \delta \dot{S^-} \rangle =&-(i\Delta_s +\Upsilon_2)\langle\delta S^- \rangle-ig(\langle\delta S^- \rangle \tau_0 +S^-_0 \langle \delta \tau \rangle) \notag\\&
-2i\Omega \langle \delta S_z \rangle -\frac{2i\mu E_2 e^{-i\delta t}} \hbar \langle\delta S_z \rangle ,
\end{align}
\begin{align}
\langle \delta {\ddot \tau} \rangle +\gamma_n \langle \delta \dot{\tau} \rangle +\omega_r^2 \langle \delta \tau \rangle =-2g \omega_r \langle \delta S_z \rangle.
\end{align}
For the solution of Eqs. (0.2)-(0.4), we make the ansatz [1.2]:
\begin{gather}
\langle \delta S_z \rangle=S_+^z e^{-i \delta t}+S_-^z e^{i \delta t},\\
\langle \delta S^- \rangle=S_+ e^{-i \delta t}+S_- e^{i \delta t},\\
\langle \delta \tau \rangle=\tau_+  e^{-i \delta t}+\tau_- e^{i \delta t}.
\end{gather}
The first order linear optical susceptibility can be described as
\begin{equation}
\chi^{(1)} =\frac{\mu S_+}{E_2}.
\end{equation}
With the ralationship of $\Upsilon_1=2\Upsilon_2$  assumed , upon substituting the above ansatz into Eqs. (0.2)-(0.4) and using Eq. (0.8), we can obtain
\begin{equation}
\chi^{(1)}=\frac{\mu^2}\hbar \frac{\omega_0 A+B^* C}{C \Omega-AD},
\end{equation}
where
\begin{align}
A&=2 g^2 \eta B^* \Omega-2\Omega^2 +E(\delta+i\Upsilon_1),\notag\\
B&=\frac{\Omega\omega_0}{i\Upsilon_2 -\Delta_s +\frac{g^2 \omega_0}{\omega_r}},\notag\\
C&=2E(g^2 \eta B -\Omega), \notag\\
D&=\Delta_s -i\Upsilon_2 -g^2 \frac{\omega_0}{\omega_r}-\delta,\notag\\
E&=\Delta_s +i\Upsilon_2 -g^2 \frac{\omega_0}{\omega_r}+\delta,\notag\\
\end{align}
the auxiliary function $\eta$   satisfies
\begin{equation}
\eta=\frac{\omega_r}{\omega_r^2 -\delta^2 -i\delta\gamma_n},
\end{equation}
and the population inversion $\omega_0$  is determined by
\begin{equation}
(\omega_0 +1)[\Upsilon_2^2 +(g^2 \frac{\omega_0}{\omega_r}-\Delta_s)^2]=-2 \Omega^2 \omega_0.
\end{equation}